# Proteins' Evolution Upon Point Mutations

Jorge A. Vila

IMASL-CONICET, Universidad Nacional de San Luis, Ejército de Los Andes 950, 5700 San Luis, Argentina.

The primary aim of this work is to explore how proteins' point mutations impact their marginal stability and, hence, their evolvability. With this purpose, we show that the use of four classic notions, namely, those from Leibniz & Kant (1768), Maynard Smith (1970), Einstein & Infeld (1961), and Anfinsen (1973), is sufficient for a better understanding of the protein-evolution and, consequently, to determine the factors that could control it. The preliminary results -without considering epistasis effects explicitly- indicate that the protein marginal-stability change upon point mutations provides the necessary and sufficient information to describe, through a Boltzmann factor, the evolution of the amide hydrogen-exchange protection factors. This finding is of paramount importance because it illustrates the impact of point mutations on both the protein marginal-stability and the ensemble of folded conformations coexisting with the native state and, in the presence of metamorphism, on the propensity for the appearance of new folds and functions.

The term epistasis has been used with various analogous meanings, although it is commonly defined as a phenomenon that "… *occurs when the combined effect of two or more mutations differs from the sum of their individual effects…*" (Milton *et al*., 2020). We should point out here a remarkable equivalence between this definition of epistasis with Leibniz & Kant's notion of space (and time) devised as "analytic wholes", *i.e*., the one where "…*its priority makes it impossible to obtain it by the additive synthesis of previously existing entities*…" (Gomez, 1998). Beyond this philosophical thought, the following question arises, why should we be interested in epistasis? The main reason is that epistasis could have a remarkable impact on the evolution of proteins by either restricting their trajectories or opening new paths to new sequences that would otherwise have been inaccessible (Miton & Tokuriki, 2016; Milton *et al*., 2020; Domingo *et al*., 2020 and reference therein). However, despite the simplicity of the definition of epistasis and the colossal progress in the prediction of protein's structures (Jumper *et al*., 2021) the answers to simple questions remain elusive because "…*there is currently no means to predict specific epistasis from a protein sequence or structure*…" (Miton *et al*., 2020). What if we analyze a far simpler problem? For example, can point mutation effects be forecast accurately? Unfortunately, the answer still is no. To determine the nature of this problem, which includes that of epistasis, we should start by identifying the molecular origin and the main factors affecting point mutations. In this regard, it is not enough to consider the protein sequence nor the mutation-types -it could also be a post-translational modification (Martin & Vila, 2020)- but a precise determination of the '*field*' between and around amino-acids. The relevance of the '*field*' for an accurate description of any physical problem was highlighted by Einstein & Infeld (1961) in the following terms: "*A new concept appears in physics, the most important invention since Newton's time: the field. It needed great scientist imagination to realize that it is not the charges nor the particles, but the field in the space between the charges and the particles which is essential for the description of physical phenomena. The field concept proves most successful and leads to the formulation of Maxwell's equations* ..." The application of this concept on structural biology started with the pioneer development of all-atom '*force-field*' (Némethy & Scheraga, 1965; Gibson & Scheraga, 1967; Scheraga, 1968; Lifson & Warshel, 1968; Momany *et al*., 1975) aimed to predict the three-dimensional structure of proteins with the only knowledge of the amino-acid sequence -the protein folding problem (Anfinsen, 1973)-. However, a definitive solution to this problem has been elusive since then, *e.g*., how a sequence encodes the protein folding remains unknown (Cramer, 2021), even though the protein's three-dimensional structure can be accurately determined (Jumper *et al*., 2021). Consequently, and beyond any doubt, an accurate determination of point mutations effects is an unsolved problem yet (Serpell *et al*., 2021). Indeed, the use of a large number of methods and approaches to predict protein stability upon point mutation, *e.g*., by using physical, statistical, or empirical '*force-field*', respectively, or machine learning methods (Khan & Vihinen, 2010; Sanavia *et al*., 2020, and references therein), show limited performance and suffer from caveats (Tokuriki *et al*., 2007; Potapov *et al*., 2009; Khan & Vihinen, 2010; Kulshreshtha *et al*., 2016; Caldararu *et al*., 2021, and reference therein). Therefore, we should focus on the global rather than on the specific mutation effects.

In contrast to the mutation-specific effects, the global distribution of the proteins' stability upon mutations (Guerois, *et al*., 2002; Schymkowitz *et al*., 2005) can be forecast with acceptable accuracy by a bi-Gaussian function (Tokuriki *et al*., 2007). Such changes in the proteins' stability due to point mutations can be determined experimentally from the unfolding Gibbs free energy ($\Delta G_U$) between the wild-type (*wt*) and the mutant (*m*) protein, *viz*., as $\Delta\Delta G_U = (\Delta G_U^m - \Delta G_U^{wt})$ (Bigman & Levy, 2018). Consequently, considering that point mutations *mainly affects* the native state stability (Zeldovich *et al*., 2007), the observed change on the unfolding Gibbs free energy ($\Delta\Delta G_U$) should represent, fundamentally, the change ($\Delta\Delta G$) in the protein marginal-stability ($\Delta G$), which refers to the Gibbs free-energy gap between the native state and the first unfolded state (Hormoz, 2013; Vila, 2019; Martin & Vila, 2020). Let us provide some pieces of evidence that support this important conjecture. The proteins' free-energy of unfolding ($\Delta G_U$) spans a wide range of variations, *viz*., between 5-25 kcal/mol (Gromiha *et al*., 2016), revealing the complexity of the 'protein folding problem' (Levinthal, 1968; Anfinsen, 1973; Anfinsen & Scheraga, 1975; Dill *et al*., 2008). However, its range of variation upon point mutations ($\Delta\Delta G_U$) is small and well-defined, revealing the validity of the thermodynamic hypothesis -Anfinsen's dogma- (Anfinsen, 1973), as explained next. The absolute value of the unfolding Gibbs free energy changes ($|\Delta\Delta G_U|$) -from the histogram of more than 5,200 point mutation data obtained by using urea and thermal unfolding experiments (Zeldovich *et al*., 2007)- are within the following narrow range of variation: $|\Delta\Delta G_U| \leq \sim 7.4$ kcal/mol. Notably, this boundary value for $|\Delta\Delta G_U|$ belongs to the proteins' marginal stability upper bound (Vila, 2019), which (*i*) is a universal feature of proteins, *i.e.*, was obtained regardless of the fold-class or its amino-acid sequence (Vila, 2019); (*ii*) is a consequence of Anfinsen's dogma validity (Vila, 2019; Vila, 2021); and (*iii*) represents a threshold beyond which a conformation will unfold and become nonfunctional (Martin & Vila, 2020; Vila, 2021). The latter means that changes in the Gibbs free-energy gap size ($\Delta\Delta G$) between the native state and the first unfolded state cannot be larger than ~7.4 kcal/mol, *e.g.*, as it occurs for the single-mutants of the green fluorescent protein from Aequorea victoria that loses ~100% of the log-fluorescence (native function) if $\Delta\Delta G_U > \sim 7.5$ kcal/mol (Sarkisyan, *et al*., 2016). Consequently, assuming $\Delta\Delta G \sim \Delta\Delta G_U$ -with the latter being experimentally determined- is a reasonable strategy to obtain a reliable assessment of the change on the protein marginal-stability upon point mutations and, from here, their effects on the ensemble of folded conformations coexisting with the native state (Vila, 2021), as shown later.

The gained knowledge on protein (*i*) stability (Albert, 1989; Tokuriki *et al*., 2007; Zeldovich *et al*. 2007; Taverna & Goldstein 2002; Williams *et al*., 2006; Vila, 2019; Martin & Vila, 2020), (*ii*) metamorphism, characterized by the existence of two or more folds with a significant structural difference between them (Luo *et al*., 2008; Murzin, 2008; Tuinstra *et al*., 2008; Alexander *et al*., 2009; Yadid *et al*. 2010; Lella *et al*., 2017; Dishman & Volkman, 2018; Shortle, 2010; Vila, 2020b; Dishman *et al*., 2021), and (*iii*) evolvability, the ability of a biological system to provide, by mutation and selection, phenotypic variation (James & Tawfik, 2003; Bloom *et al*., 2006; Romero & Arnold, 2009; Tokuriki & Tawfik, 2009; Tokuriki & Tawfik, 2009b; Breen *et al*., 2012; Goldstein, 2018; Dishman *et al*., 2021) has been enormous. This will enable us to examine below, in light of evolution, how point mutations could impact each of those issues.

The unfolding Gibbs free energy changes upon mutation ($\Delta\Delta G_U$) -in kcal/mol- instantly enables us to determine if they are positive (stabilizing) or negative (destabilizing) contributions to the protein's marginal stability. However, their impact on both the ensemble of folded conformations in equilibrium with the native-state or the metamorphism propensity cannot be straightforwardly inferred. To solve this issue, the amide Hydrogen eXchange (HX) may be used because it is a sensitive probe to assess changes in the protein native-state structure (Hvidt & Linderstrøm-Lang, 1954; Berger *et al*., 1957; Privalov &Tsalkova, 1979; Englander *et al*., 1997; Huyghues-Despointes *et al*., 1999; Craig *et al*., 2011; Balasubramaniam & Komives, 2013; Persson & Halle, 2015). Indeed, their use could bring precise information on the structural changes that could occur upon mutations and, consequently, on their impact on the protein marginal stability (Vila, 2021). This is possible because the intra- and inter-molecular hydrogen bonds are dependent on the protein native-state structure and the milieu. A simple example will be enough to illustrate this methodology. Shirley *et al*. (1991) accurately determined the urea and thermal unfolding average free-energy change ($\Delta\Delta G_U$) on ribonuclease T1 (Rnase T1) for 12-point mutations involving Tyr→Phe, Ser→Ala and Asn→Ala amino-acids, respectively. As a result, the observed destabilizing average $\Delta\Delta G_U$ values were within the following range of variation: $\approx -0.5$ kcal/mol (for Tyr57→Phe) to $\sim -2.9$ kcal/mol (for Asn81→Ala). Before we proceed, let us remember: firstly, that $\Delta\Delta G \sim \Delta\Delta G_U$ will provide us with the Gibbs free energy change in the protein's marginal stability upon point mutation; secondly, the amide HX protection factor ($P_f$) for a protein in their native-state, *i.e.*, in the EX2 limit (Bahar *et al*., 1998), is given by the following equation $\Delta G_{HX} = RT \ln P_f$ (Bahar *et al*., 1998), where $\Delta G_{HX}$ represent the Gibbs free-energy change for the opening/closing equilibrium (Bahar *et al*., 1998; Craig *et al*., 2011), $R$ is the gas constant, and $T$ the absolute temperature. Because our interest focuses on a particular region of the conformational space, namely, in the Boltzmann-ensemble of folded states in equilibrium with the native state, the following relation $\Delta G_{HX} \sim \Delta G$ should hold (Vila, 2021). Consequently, upon a point mutation the following relations $\Delta\Delta G_{HXm,wt} \sim \Delta\Delta G = (\Delta G_m - \Delta G_{wt}) = RT \ln (P_{f,m}/P_{f,wt})$, where $P_{f,m}$ and $P_{f,wt}$ are the corresponding protection factors for the mutant (*m*) and the wild-type (*wt*) protein, respectively, should also hold. Therefore, if $\Delta\Delta G \sim -2.9$ kcal/mol, then $P_{f,m} \sim P_{f,wt} \times 10^{-2}$, where $P_{f,wt}$ represents the resistance of the amide HX in the wild-type native state relative to that of the highest free-energy conformation in the ensemble of folded states (Vila, 2021). In other words, because the Asn81→Ala is a destabilizing mutation, it will leave a native state for the mutant (*m*) that is ~100 times less resistant to the amide hydrogen exchange than that of the wild-type protein (*wt*). Therefore, the point mutations not solely change the stability of the native state but also the structural dispersion in the ensemble of folded conformations coexisting with it. This conjecture is in line with convincing theoretical simulations of the hydrogen-exchange mechanism on proteins (Miller & Dill, 1995; Vendruscolo *et al*., 2003). Indeed, such simulations show that sizeable structural differences -in the ensemble of folded conformations relative to the native state- are not only likely but necessary for accurately analyzing the observed amide hydrogen exchange. From this point of view, it is reasonable to assume that a point mutation will introduce structural/energetic fluctuations in the ensemble of native folds in equilibrium with the native state. In particular, if the protein were a metamorphic one -an attribute encoded in its amino-acid sequence- a point mutation could modify, *a priori*, its metamorphism propensity. Regardless of this, the appearance of fluctuations in the ensemble of native folds could allow redistribution of their folded states ratio -determined by its Boltzmann factors- and, hence, could benefit/impair the thermodynamic equilibrium between highly dissimilar (metamorphic) folded states (Vila, 2021). This could be of paramount importance to identify critical amino-acid for the arise -or disappearance- of metamorphism in proteins such as in the study of the appearance of new folds and functions upon a mutation (Shortle, 2009; Dishman *et al*., 2021). In addition to all of the above, the protein evolvability should also be affected by point mutation because it is well-known that stability promotes it (Wagner, 2005; Bloom *et al*., 2006; Arnold, 2009; Romero & Arnold, 2009).

We are now in good condition to determine how a series of point mutations will affect the protein's marginal-stability change in light of evolution. To solve this, let us frame the problem within the Protein Space model of evolution where "…*two sequences are neighbors if one can be converted into another by a single amino-acid substitution* …" (Maynard Smith, 1970). Implicit in this modeling is that any functional protein that pertains to the Protein Space obeys Anfinsen's dogma (Vila, 2020; Vila, 2020b). Indeed, in his seminal 1970 article, John Maynard Smith (1970) envisioned that "…*if evolution by natural selection is to occur, functional proteins must form a continuous network which can be traversed by unit mutational steps without passing through nonfunctional intermediates*…" Then, a walk in that Protein Space enables us to determine, after *j* consecutive point mutations steps, the following relations of interest:

$$(\Delta G_U^1 - \Delta G_U^{wt}) + \sum_{s=2}^{j} \Delta\Delta G_U^s = (\Delta G_U^j - \Delta G_U^{wt}) = \Delta\Delta G_U^j \tag{1}$$

where $\Delta\Delta G_U^s = (\Delta G_U^s - \Delta G_U^{s-1})$ for *s* > 1, and *wt* for the wild-type. Then, considering

$$(\Delta G^1 - \Delta G^{wt}) + \sum_{s=2}^{j} \Delta\Delta G^s = (\Delta G^j - \Delta G^{wt}) = \Delta\Delta G^j = RT \ln\left(\frac{P_{f,j}}{P_{f,wt}}\right), \text{ and } \Delta\Delta G^j \sim \Delta\Delta G_U^j \tag{2}$$

the following relationship, in terms of the observable $\Delta\Delta G_U^j$, can be obtained:

$$P_{f,j} \sim P_{f,wt}\, e^{\beta \Delta\Delta G_U^j} \tag{3}$$

with $\beta = 1/RT$ and, $P_{f,j}$ and $P_{f,wt}$ are the corresponding protection factors for the protein after *j* point mutations steps and the wild-type (*wt*) native state, respectively. In other words, Eq. (3) represents -after navigating the Protein Space as an abstract model of evolution- the changes on both the wild-type protein native-state stability ($\Delta\Delta G_U^j$) and the ensemble of folded conformations coexisting with it ($P_{f,j}$). Worth is noting that the results in Eq. (1) to (3) are valid even if there are *k* out of *j* mutations (with *k* < *j-1*) leading to nonfunctional proteins, *e.g.*, when a mutation leads to a free-energy change (ΔΔG) larger than the marginal stability upper bound threshold of ~7.4 kcal/mol (Vila, 2019; Vila 2021). Consideration of this problem is relevant for two reasons. Firstly, because most of the point mutations are destabilizing (Zeldovich et al., 2007; Tokuriki & Tawfik, 2009;

Arnold, 2009; Socha & Tokuriki, 2013) and, secondly because the evolutionary trajectories in the protein sequence space are assumed to be reversible although the genotypic irreversibly should not be dismissed (Kaltenbach *et al*., 2015).

Let us assume that the ancestor and target protein sequence, respectively, are kept fixed during the evolutive process, as in the word-game (Maynard Smith, 1970). In this game, one-word changes into another by replacing one letter at a time, *e.g*., transforming the word 'NCSI' to 'IRNL' (Ogbunugafor, 2020), where each letter represents an amino-acid in the single-letter code. Then, according to Eq. (3), turning 'NCSI' into 'IRNL' will lead to the same result for $P_{f,j}$, whatever the number (*j*) of point mutations steps used. That is feasible because $\Delta\Delta G^j$ ($\sim\Delta\Delta G_U^j$) is a state function. Thus, nature follows any evolutive path if there is no penalty for doing so. The word-game also enables us to rationalize the complexity of the protein evolution analysis, *e.g*., in terms of either evolutionary trajectories' or marginal stability changes. Let us briefly discuss the pros and cons of each of these approaches.

On the one hand, an analysis of the protein evolution in terms of the evolutionary trajectories implies knowing with precision, in each step, the letter to be mutated (amino-acid identity), the background where the mutation occurs (sequence and milieu), and the epistasis effects that could take place. An accurate solution to this problem is a daunting task (Sailer & Harms, 2017), although of great practical relevance, *e.g*., how to turn a protein to exhibit the desired function -as it happens on directed evolution applications (Arnold, 2009; Romero & Arnold, 2009). A solution to this particular problem is exacerbated by the fact that neutral mutations, asides from epistasis effects, also need to be considered because they may play a critical role in the transition from one amino-acid to another (Kimura, 1968; Wagner, 2005; Bloom et al., 2006; Bloom & Arnold, 2009; Draghi *et al*., 2010). In other words, neutral mutations (which are invisible to natural selection) may compensate for the effects of destabilizing mutations though beneficial from the functional point of view (Bloom & Arnold; Arnold, 2009; Draghi *et al*., 2010).

On the other hand, if protein stability is an essential factor for evolution (Socha & Tokuriki, 2013; Kurahashi *et al*., 2018), then it is reasonable to think about the whole (marginal-stability evolution) rather than the parts (evolutionary-trajectories) because the former involves the latter and, consequently, *all* factors affecting protein evolution -including, specific, nonspecific, or 'high-order' epistasis effects (Starr & Thornton, 2016; Miton & Tokuriki, 2016; Sailer & Harms, 2017; Domingo *et al*., 2019)-. Yet, we should keep in mind that Leibniz & Kant's notion of space (and time), devised as '*analytic wholes*' (Gomez, 1998), highlights that the whole is more than the sum of the parts, although this does not imply the irrelevance of the latter. Indeed, as noted above for the directed evolution applications, if our interest focuses on predicting the evolutionary trajectories, detailed consideration of the epistasis effects should be unavoidable.

The above analysis shows that the marginal-stability is the main factor controlling protein evolution. Indeed, changes on the protein marginal-stability enabled us to quantify the fluctuations that may occur, in light of evolution, in the stability of the native-state, and on the ensemble of folded states coexistent with it and, hence, their possible impact on the progress toward new architectures and functions. This conclusion stresses the idea that proteins evolution should be primarily analyzed in terms of the "marginal-stability" rather than focus on the "trajectories", in line with Leibniz & Kant's notion of space (and time) devised as "analytic wholes".

**Acknowledgments**

I thank Marcela Renée Becerra Batán for introducing me to the reference of Ricardo J. Gomez on "Leibniz's Spark of Kant's Great Light".

I'm honored to dedicate this work to the memory of Harold A. Scheraga, a worldwide respected scientist in the biophysical chemistry field that was my mentor, colleague, and friend for more than 30 years.

The author acknowledges support from the IMASL-CONICET (PIP-0087) and ANPCyT (PICT-02212), Argentina, and the access to books and journals, through the Mann, Olin and Uris libraries of Cornell University.

Corresponding author
Jorge A. Vila jv84@cornell.edu
https://orcid.org/0000-0001-7557-9350

**References**

Albert, T. Mutational effects on protein stability. *Annu. Rev. Biochem*. **1989**, 58, 765-798.

Alexander, P.A.; He, Y.; Chen, Y.; Orban, J.; Bryan, P.N. A minimal sequence code for switching protein structure and function. *Proc Natl Acad Sci USA*. **2009,** 106, 21149-21154.

Anfinsen, C.B. Principles that govern the folding of protein chains. *Science*. **1973**, 181, 223-230.

Anfinsen, C.B.; Scheraga, H.A. Experimental and Theoretical Aspects of Protein Folding. *Advances in Protein Chemistry*. **1975**, 29, 205-300.

Arnold, F.H. How proteins adapt: lessons from directed evolution. *Cold Spring Harbor Symposia Quantitative Biology*. **2009**, 74, 41-46.

Bahar, I.; Wallqvist, A.; Covell, D.G.; Jernigan, R.L. Correlation between native-state hydrogen exchange and cooperative residue fluctuation from a simple model. *Biochemistry*. **1998**, 37, 4, 1067-1075.

Balasubramaniam, D.; Komives, E.A. Hydrogen-exchange mass spectrometry for the study of intrinsic disorder in proteins. *Biochim. Biophys. Acta*. **2013**, 1834, 1202-1209.

Berger, A.; Linderstrøm-Lang, K. Deuterium exchange of poly-DL-alanine in aqueous solution. *Arch. Biochem. Biophys*. **1957**, 69, 106-118.

Bigman, L.S.; Levy, Y. Stability Effects of Protein Mutations: The Role of Long-Range Contacts. *The Journal of Physical Chemistry B*. **2018**, 122, 11450-11459.

Bloom, J.D.; Arnold, F.H. In the light of directed evolution: pathways of adaptive protein evolution. *Proc Natl Acad Sci U S A*. **2009**, 106, 9995-10000.

Bloom, J.D.; Labthavikul S.T.; Otey C.R., Arnold, F.H. Protein stability promotes evolvability. *Proc Natl Acad Sci USA*. **2006**,103:5869–5874.

Breen, M.; Kemena, C.; Vlasov, P. *et al.* Epistasis as the primary factor in molecular evolution. *Nature* **490,** 535–538 (2012).

Caldararu, O.; Blundell, T.L; Kepp, K.P. A base measure of precision for protein stability predictors: structural sensitivity. *BMC Bioinformatics.* **2021**, 22**:**88, 2-14.

Craig, P.O.; Lätzer, J.; Weinkam, P.; Hoffman, R.M.B.; Ferreiro, D.U.; Komives, E.A.; Wolynes, P.G. Prediction of Native-State Hydrogen Exchange from Perfectly Funneled Energy Landscapes. *J. Am. Chem. Soc*. **2011**, 133, 17463-17472.

Cramer, P. AlphaFold2 and the future of structural biology. *Nat Struct Mol Biol*. **2021**, 28, 704-705.

Dill, K.A.; Ozkan, B.S.; Shell, M.S.; Weikl, T.R. The protein folding problem. *Annual Review of Biophysics*. **2008** 37:1, 289-316

Dishman, A.F.; Tyler, R.C.; Fox, J.; Kleist, A.B.; Prehoda, K.E.; Babu, M.M.; Peterson, F.C.; Volkman, B.F. Evolution of fold switching in a metamorphic protein. *Science.* **2021***,* 371, 86-90.

Dishman, A.F.; Volkman, B.F. Unfolding the Mysteries of Protein Metamorphosis. *ACS Chemical Biology*. **2018**, 13, 1438-1446.

Domingo, J.; Baeza-Centurion, P.; Lehner, B. The Causes and Consequences of Genetic Interactions (Epistasis). *Annual Review of Genomics and Human Genetics*. **2019**, 20, 433-460.

Draghi, J.A.; Parsons, T.L.; Wagner, G.P.; Plotkin, J.B. Mutational robustness can facilitate adaptation. *Nature*. **2010**, 463, 353-555.

Einstein, A.; Infeld, L. The Evolution of Physics. **1961**, Published by Simon and Schuster, Inc., New York, pp. 244.

Englander, S.W.; Mayne, L.; Bai Y.; Sosnick, T.R. Hydrogen exchange: the modern legacy of Linderstrøm-Lang. *Protein Sci*. **1997**, 6, 1101-1109.

Gibson, K.D.; Scheraga, H.A. Minimization of polypeptide energy. I. Preliminary structures of bovine pancreatic ribonuclease S-peptide, *Proc Natl Acad Sci USA*. **1967**, 58, 420-427.

Goldstein, R.A. The evolution and evolutionary consequences of marginal thermostability in proteins. *Proteins*. **2018**, 79, 1396-1407.

Gomez, R.J. Leibniz's Spark of Kant's Great Light, chapter 14 in 'Thought, Language, and Ontology'. **1998,** Orilia F. and Rapaport W., eds. Kluwer Academic Publishers, Netherlands, pp 313-329.

Gromiha, M.M.; Anoosha, P.; Huang, L.T. Applications of Protein Thermodynamic Database for Understanding Protein Mutant Stability and Designing Stable Mutants. *Methods Mol Biol*. **2016**, 1415, 71-89.

Guerois, R.; Nielsen, J.E.; Serrano, L. Predicting Changes in the Stability of Proteins and Protein Complexes: A Study of More Than 1000 Mutations. *Journal of Molecular Biology*. **2002**, 320, 369-387.

Hormoz, S. Amino acid composition of proteins reduces deleterious impact of mutations. *Sci Rep*. **2013**, 3, 1-10.

Huyghues-Despointes, B.; Scholtz, J.; Pace, C. Protein conformational stabilities can be determined from hydrogen exchange rates. *Nat Struct Mol Biol*. **1999**, 6, 910-912.

Hvidt, A.; Linderstrøm-Lang, K. Exchange of hydrogen atoms in insulin with deuterium atoms in aqueous solutions. *Biochim Biophys Acta*. **1954**, 14, 574-575.

James, L.C.; Tawfik, D.S. Conformational diversity and protein evolution -a 60-year-old hypothesis revisited. *Trends Biochem Sci*. **2003**, 28, 361-368.

Jumper *et al*. Highly accurate protein structure prediction with AlphaFold. *Nature*. **2021**, 596, 583-589.

Kaltenbach, M.; Jackson, C.J.; Campbell, E.C.; Hollfelder, F.; Tokuriki, N. Reverse evolution leads to genotypic incompatibility despite functional and active site convergence. *Elife*. **2015**, 4:e06492.

Khan, S.; Vihinen, M. Performance of protein stability predictors. *Human Mutation*. **2010**, 31, 675-684.

Kimura, M. Evolutionary rate at the molecular level. *Nature*. **1968**, 217, 624-626.

Kulshreshtha, S.; Chaudhary, V.; Goswami, G.K.; Mathur, N. Computational approaches for predicting mutant protein stability. *J Comput Aided Mol Des*. **2016**, 30, 401-12.

Kurahashi, R.; Sano, S.; Takano, K. Protein Evolution is Potentially Governed by Protein Stability: Directed Evolution of an Esterase from the Hyperthermophilic Archaeon Sulfolobus tokodaii. *J Mol Evol*. **2018**, 86, 283-292.

Lella, M.; Mahalakshmi, R. Metamorphic Proteins: Emergence of Dual Protein Folds from One Primary Sequence. *Biochemistry*. **2017**, 56, 2971-2984.

Levinthal, C. Are There Pathways for Protein Folding? *Journal de Chimie Physique*. **1968**, 65:44-45.

Lifson, S.; Warshel, A. Consistent Force Field for Calculations of Conformations, Vibrational Spectra, and Enthalpies of Cycloalkane and *n*-Alkane Molecules. *The Journal of Chemical Physics*. **1968**, 49, 5116-5129.

Luo, X.; Yu, H. Protein Metamorphosis: The two-state behavior of Mad2. *Structure*. **2008**, 16, 1616-1625.

Markin, C.J.; Mokhtari, D. A.; Sunden, F.; Appel, MJ.; Akiva, E.; Longwell, S.A.; Sabatti, C.; Herschlag, D.; Fordyce, P.M. Revealing enzyme functional architecture via high-throughput microfluidic enzyme kinetics. *Science.* **2021**, 373, 6553.

Martin, A.O.; Vila, J.A. The Marginal Stability of Proteins: How the Jiggling and Wiggling of Atoms is Connected to Neutral Evolution. *Journal Molecular Evolution*. **2020**, 88, 424-426.

Maynard Smith, J. Natural Selection and the concept of a protein space. *Nature*. **1970**, 225, 563-564.

Miller, D.W.; Dill, K.A. A statistical mechanical model for hydrogen exchange in globular proteins. *Protein Sci*. **1995**, 4, 1860-1873.

Miton, C.M.; Buda, K.; Tokuriki, N. Epistasis and intramolecular networks in protein evolution. *Current Opinion in Structural Biology*. **2021**, 69, 160-168.

Miton, C.M.; Chen, J.Z.; Ost, K.; Anderson, D.W.; Tokuriki, N. Statistical analysis of mutational epistasis to reveal intramolecular interaction networks in proteins. *Methods Enzymol.* **2020**, 643, 243-280.

Miton, C.M.; Tokuriki, N. How mutational epistasis impairs predictability in protein evolution and design. *Protein Sci*. **2016**, 25, 1260-1272.

Momany, F.A.; McGuire, R.F.; Burgess, A.W.; Scheraga, H.A. Energy parameters in polypeptides. VII. Geometric parameters, partial atomic charges, nonbonded interactions, hydrogen bond interactions, and intrinsic torsional potentials for the naturally occurring amino acids. *J Phys Chem*. **1975**, 79, 2361-2381.

Murzin, A.G. Biochemistry. Metamorphic proteins. *Science*. **2008**, 320, 1725-1726.

Némethy, G.; Scheraga, H.A. Theoretical determination of sterically allowed conformations of a polypeptide chain by a computer method, *Biopolymers*. **1965**, 3, 155-184.

Ogbunugafor, CB. A Reflection on 50 Years of John Maynard Smith's "Protein Space". *Genetics*. **2020**, 214, 749-754.

Persson, F.; Halle, B. How amide hydrogens exchange in native proteins. *Proc Natl Acad Sci USA*. **2015**, 112, 10383-10388.

Potapov, V.; Cohen, M.; Schreiber, G. Assessing computational methods for predicting protein stability upon mutation: good on average but not in the details, *Protein Engineering, Design and Selection*. **2009**, 22, 553-560.

Privalov, P.L.; Tsalkova, T.N. Micro- and macro-stabilities of globular proteins. *Nature*. **1979**, 280, 693-696.

Romero, P.A.; Arnold, F.H. Exploring protein fitness landscapes by directed evolution. *Nat Rev Mol Cell Biol*. **2009**, 10, 866-876.

Sailer, ZR; Harms, MJ. Molecular ensembles make evolution unpredictable. *Proc Natl Acad Sci USA*. **2017**, 114,11938-11943.

Sanavia, T.; Birolo, G.; Montanucci, L.; Turina, P.; Capriotti, E.; Fariselli, P. Limitations and challenges in protein stability prediction upon genome variations: towards future applications in precision medicine. *Comput Struct Biotechnol J*. **2020**, 18, 1968-1979.

Sarkisyan, K.S.; Bolotin, D.A.; Meer, M.V.; *et al*. Local fitness landscape of the green fluorescent protein. *Nature*. **2016**, 533, 397-401.

Scheraga, HA. Calculations of conformations of polypeptides. *Adv Phys Org Chem*. **1968**, 6, 103-184.

Schymkowitz, J.W.; Rousseau, F.; Martins, I.C.; Ferkinghoff-Borg, J.; Stricher, F.; Serrano, L. Prediction of water and metal binding sites and their affinities by using the Fold-X force field. *Proc Natl Acad Sci USA*. **2005**, 102, 10147-10152.

Serpell, L.C.; Radford, S.E.; Otzen, D.E. AlphaFold: A Special Issue and A Special Time for Protein Science. *J Mol Biol*. **2021**, 433, 167231.

Shirley, B.A.; Stanssens, P.; Hahn, U.; Pace, C.N. Contribution of hydrogen bonding to the conformational stability of ribonuclease T1. *Biochemistry*. **1992**, 31, 3, 725-732.

Shortle, D. One sequence plus one mutation equals two folds. *Proc. Natl. Acad. Sci. USA*. **2009**, 106, 21011-21012.

Socha, R.D.; Tokuriki, N. Modulating protein stability - directed evolution strategies for improved protein function. *FEBS J*. **2013**, 280:5582-5595.

Starr, T.N.; Thornton, J.W. Epistasis in protein evolution. *Protein Sci*. **2016**, 25, 1204-1218.

Taverna, D.M.; Goldstein, R.A. Why Are Proteins Marginally Stable? *Proteins.* **2002***,* 46, 105-109.

Tokuriki, N.; Stricher, F.; Schymkowitz, J.; Serrano, L.; Tawfik, D.S. The Stability Effects of Protein Mutations Appear to be Universally Distributed. *Journal of Molecular Biology*. **2007**, 369, 1318-1332.

Tokuriki, N.; Tawfik, D.S. Protein dynamics and evolvability. *Science*. **2009**, 324, 203-207.

Tokuriki, N.; Tawfik, D.S. Stability effects of mutations and protein evolvability. *Current Opinion in Structural Biology*. **2009b** 19, 596-604.

Tuinstra, R.L.; Peterson, F.C.; Kutlesa, S.; Elgin, E.S.; Kron, M.A.; Volkman, B.F. Interconversion between two unrelated protein folds in the lymphotactin native state. *Proc Natl Acad Sci USA*. **2008**, 105, 5057-5062.

Vendruscolo, M; Paci, E.; Dobson, C.M.; Karplus, M. Rare Fluctuations of Native Proteins Sampled by Equilibrium Hydrogen Exchange. *J. Am. Chem. Soc*. **2003**, 125, 51, 15686-15687.

Vila, J.A. About the Protein Space Vastness. *Protein J.* **2020**, 39, 472-475.

Vila, J.A. Forecasting the upper bound free energy difference between protein native-like structures. *Physica A*. **2019**, 533, 122053.

Vila, J.A. Metamorphic proteins in light of Anfinsen's dogma. *J Phys Chem Lett*. **2020b**, 11, 4998-4999.

Vila, J.A. Thoughts on the protein's native state. *J Phys Chem Lett*. **2021**, 12, 5963-5966.

Wagner, A. Robustness, evolvability, and neutrality. *FEBS Lett*. **2005**, 579, 1772-1778.

Williams, P.D.; Pollock, D.D.; Goldstein, R.A. Functionality and the evolution of marginal stability in proteins: Inferences from lattice simulations. *Evolutionary Bioinformatics Online*. **2006**, 2, 91-101.

Yadid, I.; Kirshenbaum, N.; Sharon, M.; Dym, O.; Tawfik, D.S. Metamorphic proteins mediate evolutionary transitions of structure. *Proc Natl Acad Sci USA*. **2010**, 107, 7287-7292.

Zeldovich, K.B.; Chen, P.; Shakhnovich, E.I. Protein stability imposes limits on organism complexity and speed of molecular evolution. *Proc Natl Acad Sci USA*. **2007**, 104, 16152-16157.